# New Design of Potentially Low-cost Solar Cells Using TiO$_2$/Graphite Composite as Photon Absorber

Dui Yanto Rahman[1], Mamat Rokhmat[1], Elfi Yuliza[1], Euis Sustini[1], and Mikrajuddin Abdullah[1,2a]

[1]Department of Physics, Bandung Institute of Technology

Jalan Ganeca 10, Bandung 40132, Indonesia

[2]Bandung Innovation Center

Jalan Sembrani 19, Bandung 40293, Indonesia

[a]Corresponding author email: din@fi.itb.ac.id

Abstract

We propose a solar cell design using the combination of titanium dioxide (TiO$_2$) and graphite as active photon absorbing materials. TiO$_2$ absorbs photons of nearly ultraviolet wavelengths to produce electron–hole pairs, while graphite is expected to absorb photons of longer wavelengths. Although many authors have claimed that graphite is a semimetal, we observed that a model of a solar cell containing $TiO_2$ only as the active material behaves exactly the same as a model containing graphite only as the active material. Additionally, we observed that a model of a solar cell made using a composite of $TiO_2$ and graphite as the active material had much higher efficiency than solar cells made using $TiO_2$- or graphite-only active materials. Although the highest efficiency we report here is approximately 1%, our proposed solar cell structure is promising for mass application, especially in low-income settings, owing to its easy and flexible fabrication, and easy large-scale application.

## Introduction

The use of renewable sources of energy and the search for inexpensive and easily fabricated solar cells are becoming increasingly important for countries to meet their rising energy needs. Presently available solar cells are expensive, especially for low-income communities. These communities, some of which are still living without electricity, are widespread in developing countries; however, only a few developed countries have expertise in solar-cell technologies. The users of solar cells will likely rely on these supplier countries for many decades.

Solar energy is considered a type of free energy in nature as it freely available over the Earth's surface. The ability to harvest this energy should thus be available every human being. This objective will be realized when more countries are able to produce their own solar cells. Globally, countries should therefore strive to produce practical solar cells, even if these cells have low efficiency.

In recent years, many attempts have been made to produce solar cells using cheaper materials, and easily applied and readily scalable methods [1–5]. Although the efficiency of our solar cells does not match the efficiency of presently applied solar cells, our solar cells might compete with presently applied solar cells economically, especially when considering factors such as production and maintenance costs and efficiency. Although the application of titanium dioxide ($TiO_2$) in solar cells is not new, the principle and implementation of its application in presently applied solar cells and in our proposed model greatly differ. The commonly used structure is a dye-sensitized solar cell [6], with $TiO_2$ used as a conductive medium for the transport of electrons from the dye to the electrode. In the proposed structure, $TiO_2$ particles play the role of a photon absorber and the produced electrons are transported to the electrode through metallic bridges.

We used $TiO_2$ as an active material that absorbs photons and converts them into electric current. This material has a band gap around 3.2–3.8 eV, allowing the effective absorption of ultraviolet light. Only a few electron–hole pairs are produced when the material is illuminated by the solar spectrum. To enlarge the absorption band of $TiO_2$, the doping of pure $TiO_2$ with V, Cr, Mn, Fe, Ni, or Au [7,8] and Pt, Rh, Au, Pd, Ni, or Ir [9] has been reported. The cited studies claim that the band gap of $TiO_2$ is reduced by doping, resulting in an increase in visible-spectrum absorption and, thus, higher solar cell performance.

Instead of doping pure $TiO_2$, we have tried using impure (low-grade) $TiO_2$, which is found in nature [10]. Previously, we have demonstrated that impure $TiO_2$ has a wider absorption spectrum [1–4,9] than pure $TiO_2$ and have applied the material in developing solar cells [1–5].

In addition to the absorption band, the lifetime of electron–hole pairs produced in the $TiO_2$ must be considered because it directly determines the photon-to-current conversion. The lifetime of an electron–hole pair in $TiO_2$ is very short and potentially limits the efficiency of photon-to-current conversion. The electron and hole in anatase

TiO$_2$ undergo non-exponential decay, where the hole population decays within in a few nanoseconds and the electron population persists for periods up to microseconds [11].

In previous works we deposited copper particles in the space between TiO$_2$ particles to suppress electron–hole recombination and observed greatly improved efficiency [1–5]. The formation of metal (Cu) and semiconductor (TiO$_2$) produces a high Schottky barrier that facilitates electron capture by the metal. Vorontsov reported that Pt-TiO$_2$ has the highest Schottky barrier and a longer electron–hole separation time [12,13]. Similarly, the Cu-TiO$_2$ contact in the present work might lengthen the lifetime of electron–hole pair separation and allow enough time for electrons to migrate into the metal and transport through the copper bridge. Hoffmann et al. stated that although the electron–hole pair lifetime is only a few nanoseconds, it is long enough for electrons to migrate to the TiO$_2$ surface and then be captured by the metal [14].

The conduction band and valence band of TiO$_2$ are produced by different orbitals. The valence band is developed by the d-orbital while the conduction band is developed by the s-p hybridized orbital [15]. This leads to different parity of the electron and hole pair, which suppresses the recombination probability.

In the present work, we propose a method of further improving the efficiency of TiO$_2$-based solar cells by adding particles capable of absorbing visible light. A mixture containing such particles and TiO$_2$ particles is expected to absorb more photons, with wavelengths approximately ranging from the ultraviolet to the red region. Previous investigations aimed to devise a strategy for improving solar cell efficiency using materials with different absorption bands. Kim et al. added a low-band-gap material to a main absorbing material to improve the efficiency of a plastic-based solar cell [16]. Peet et al. developed a tandem solar cell by combining the functions of low- and high-band-gap materials to widen the absorption range and observed improved efficiency [17]. The important difference between the previous method and the method reported here is the use of a cheap material and simple processing.

In the present work, we mix graphite particles with TiO$_2$ particles. The use of graphite as a conducting and catalytic counter electrode for dye-sensitized solar cells has improved efficiency [18]. However, to our knowledge, the use of graphite as a photon-absorbing material in a solar cell has not been investigated. Graphite is a semimetal with a valence and conduction band that overlap by around 0.03 eV [19,20].

## Method

The structure of the solar cell reported here is schematically shown in Fig. 1. The solar cell is composed of fluorine-doped tin oxide (FTO) as a transparent electrode, photon-absorbing particles, a polymer electrolyte and a counter electrode. We trialed three composites as photon-absorbing particles: $TiO_2$ only, graphite only, and a mixture of graphite and $TiO_2$. We also deposited a copper bridge in the space between particles to facilitate electron transport [1–5].

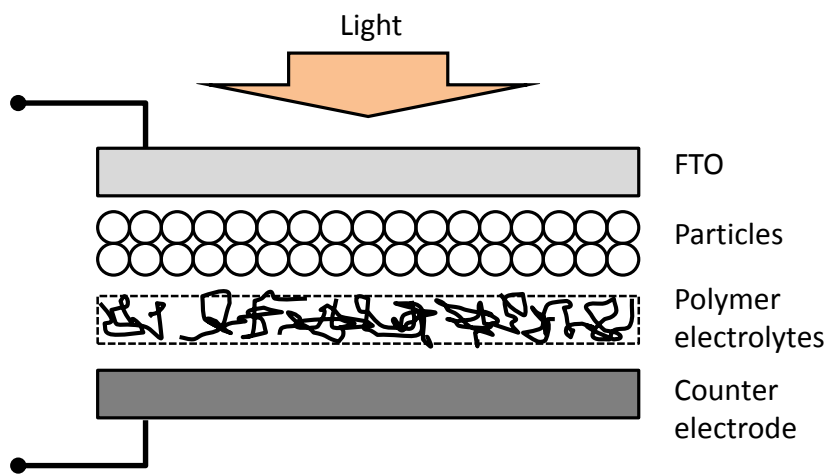

Figure 1. Schematic of solar cells investigated in this work. The solar cell is composed of a transparent conductive electrode (Fluorine-doped Tin Oxide), photon absorber particles, polymer electrolyte, and a counter electrode.

First, we made suspensions of $TiO_2$ only, graphite only, and $TiO_2$/graphite mixtures with different weight ratios in distilled water. The powders were homogeneously stirred in distilled water for 1 hour. The homogeneous suspension was then sprayed (30 times) onto the FTO surface, and heated at 200 °C to produce a thick film. The film was then heated at temperatures of 100, 200 and 300 °C for 20 minutes to evaporate water and to create better contact between the film and FTO surface.

We deposited copper in the space between particles by electroplating. Electroplating was conducted in a bath containing electrolyte liquid. The temperature of the bath was 55 °C. The electrolyte liquid was made by dissolving 14.999 g of $CuSO_4$ into 120 ml of distilled water. A copper rod (99.99% purity) was used as an anode, and the composite film as a cathode. The deposition time was controlled to be 10 s, the voltage was 5 V DC, and the separation of the anode and cathode was 2 cm.

A polymer electrolyte was made by dissolving 0.18 g LiOH (Kanto, Japan) in a beaker containing 10 ml distilled water and by dissolving 1.8 g polyvinyl alcohol (PVA) (Bratachem, Indonesia) in another beaker containing 20 ml distilled water. Each beaker was then stirred for 1 hour. Subsequently, the LiOH solution was mixed with the PVA solution and stirred for 3 hours at 100 °C to produce a gel-like electrolyte. The electrolyte was smeared manually on the film surface. Aluminum, used as a counter electrode, was then affixed to the opposite side of the electrolyte.

The morphology was observed using a scanning electron microscope (SEM) (JEOL JSM-6360LA, operated at 20 kV). A current–voltage (IV) meter (Keithley 617) was used to measure the performance of the solar cell. The measurements were made under illumination The measurements were set up under XENOPHOT 64653 HLX ELC 24V, 250W GX5.3 OSRAM lamp illumination with light intensity of 900 lumens (45 W/m$^2$) and the light intensity was measured using a lux meter.

## Results and Discussion

We first analyzed the IV behavior of solar cells containing TiO$_2$ only, graphite only, and a mixture of TiO2/graphite as the photon-absorbing material. All samples contained deposited Cu as an electron bridge to the FTO electrode. It is interesting to observe that the IV behavior of the solar cell containing graphite as the active material was similar to that of the solar cell containing TiO$_2$ as the active material. The efficiency of the solar cell containing TiO$_2$ was only 0.03% while that of the solar cell containing graphite was a much higher 0.12%. A further increase in efficiency was achieved with a mixture of TiO$_2$ and graphite as the active material. Using a weight ratio of 8% of graphite relative to TiO$_2$, efficiency of 0.27% was achieved.

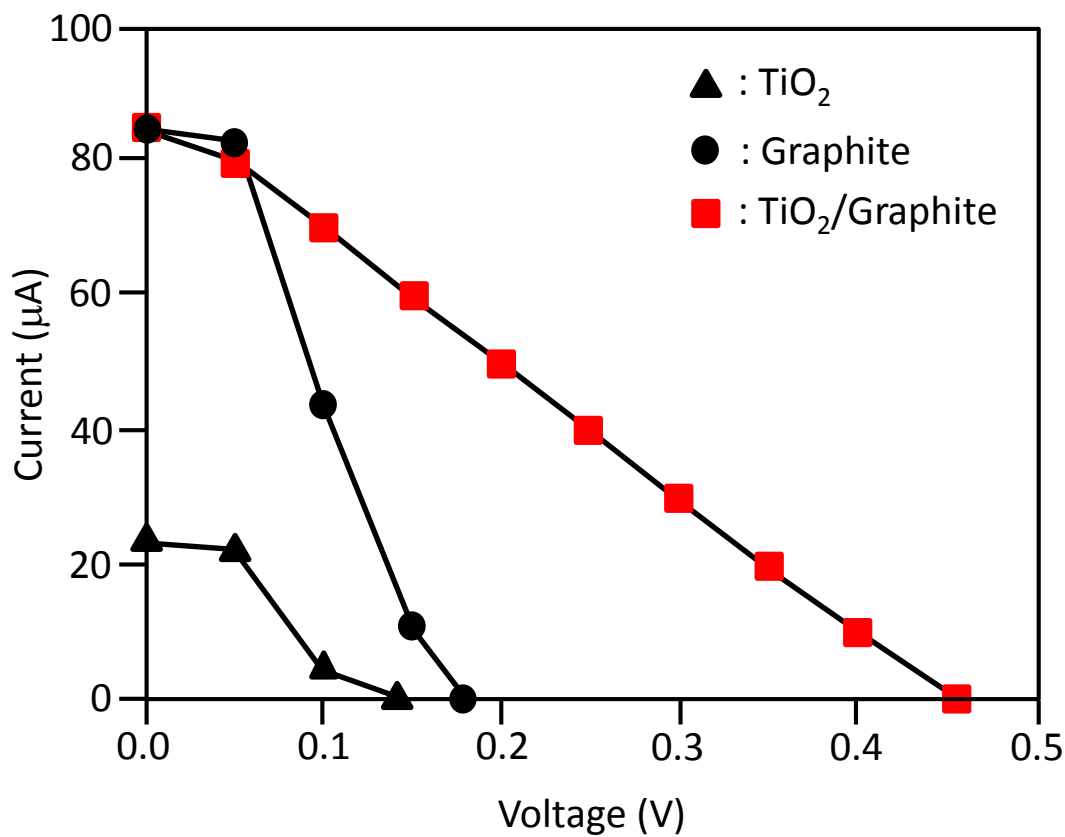

Figure 2 Current–voltage characteristics of solar cells containing an active layer of $TiO_2$ only (triangle), graphite only (circle) and a mixture of $TiO_2$/graphite (square) with a graphite/$TiO_2$ ratio of 8 %w/w. The fill factors and efficiencies of the solar cells are 0.33 and 0.03%, 0.29 and 0.12%, and 0.26 and 0.27% for cells containing $TiO_2$ only, graphite only, and a mixture of $TiO_2$/graphite, respectively.

It is clear that the mixture of $TiO_2$ and graphite achieved higher efficiency. However, the maximum observed efficiency of 0.27% remains low. Achieving a much higher efficiency remains a challenge. To this end, we investigated the deposition of multiple absorbing layers. Our basic assumption was that some of the radiation penetrating the solar cell was not converted into electricity. Thickening the active layer may maximize the photon absorption and thus improve the efficiency of conversion. We used an active layer of the $TiO_2$/graphite mixture with a weight ratio (graphite/$TiO_2$) of 8 %w/w. We deposited one, two, and three layers. Nair et al. confirmed the effects of film thickness on the structural, optical, and luminescence properties of $TiO_2$ prepared by radio frequency magnetron sputtering [21]. We expect that the film thickness also affects the efficiency of the solar cell.

Figure 3 clearly shows that, compared with the case for one layer, two active layers provide higher efficiency at 1.09%, with a fill factor of 0.28. One active layer is unable to absorb all photons, with some of the light penetrating the cells not being used to generate current. The realized efficiency was thus only 0.35%. Using two layers maximizes photon absorption and achieves a greater light-to-current conversion (efficiency). Further increasing the number of layers reduces the efficiency; the efficiency was only 0.29% for three layers. It is likely that the third layer did not receive any photons, with all photons being absorbed by the first and second layers. The creation of electrons and holes thus occurs optimally in the first and second layers. The number of electrons produced by the solar cell is nearly the same for two layers and for more than two active layers. However, the use of more than two layers adds more space for electrons to spread out and some move away from the electrode. This electron will have enough time to recombine again with a hole and does not take part in the generation of current. Consequently, the efficiency of the solar cell is reduced.

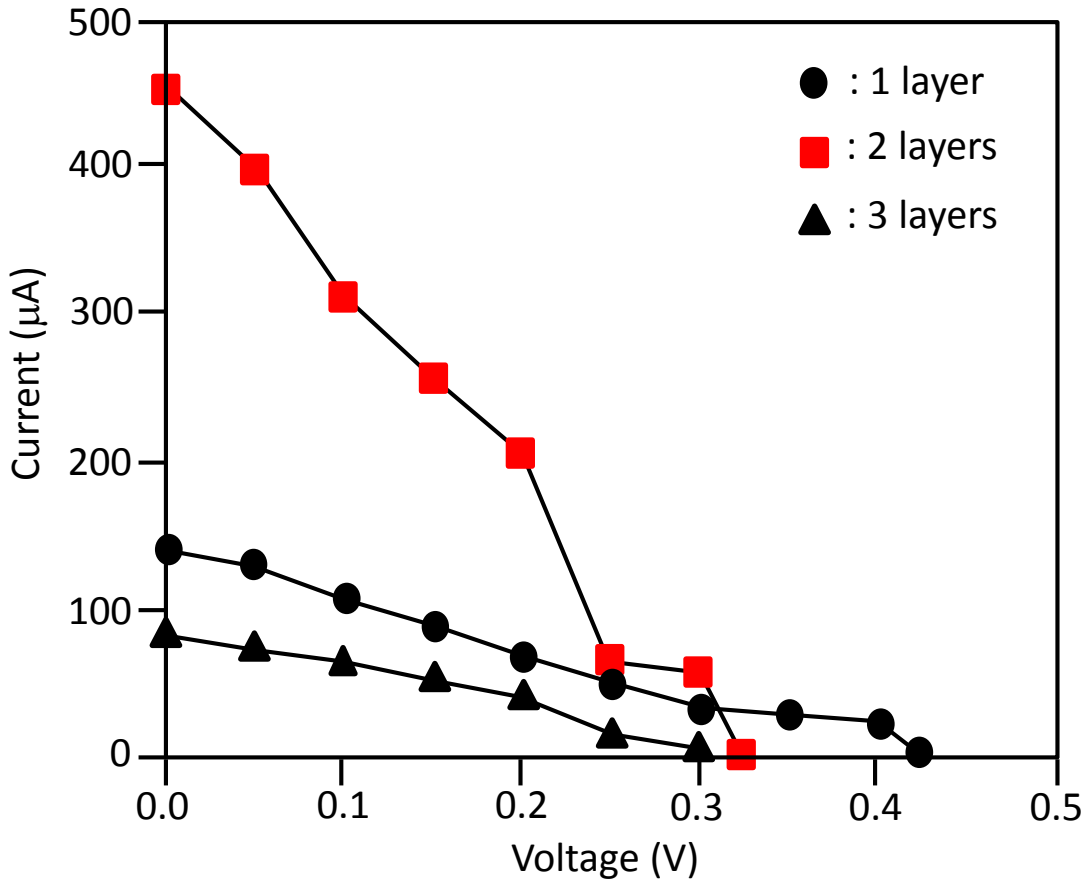

Figure 3 Effect of the number of active layers on the current–voltage characteristic of the solar cells: (circle) one layer, (square) two layers, and (triangle) three layers. An active layer is composed of $TiO_2$ and graphite with Cu particles deposited in the space between $TiO_2$ and graphite particles. The measurement was made under a xenon lamp with illumination intensity of 37.38 W/m$^2$.

An SEM image of a graphite film sprayed onto an FTO substrate is shown in Fig. 4(a). The particles are of micrometer size. An SEM image of the graphite/$TiO_2$ composite inserted with particles copper is shown in Fig. 4(b). The average size of a $TiO_2$ particle is approximately 160 nm, while the copper particles are larger than the $TiO_2$ particles.

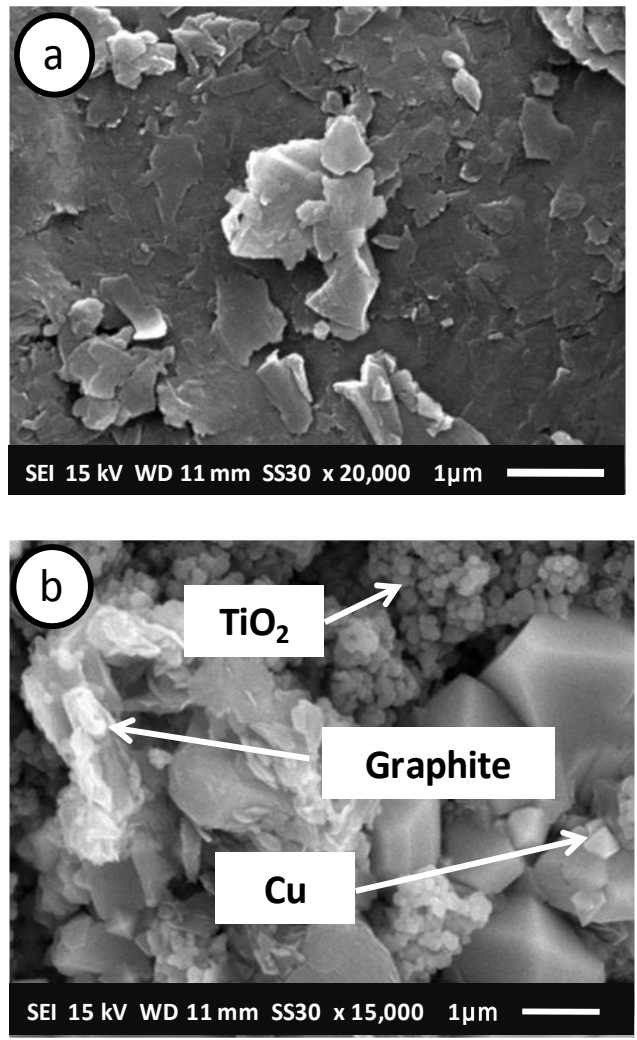

Figure 4. SEM images of (a) graphite and (b) Cu/graphite/$TiO_2$ film

The ultraviolet–visible light characterization of the composite graphite/$TiO_2$ film is presented in Fig. 5. The graphite/$TiO_2$ film absorbs a wide range of the solar spectrum, from less than 400 nm to more than 850 nm. The film absorbs a wider solar spectrum than pure $TiO_2$ does. Pure $TiO_2$ only absorbs the solar spectrum below 400 nm.

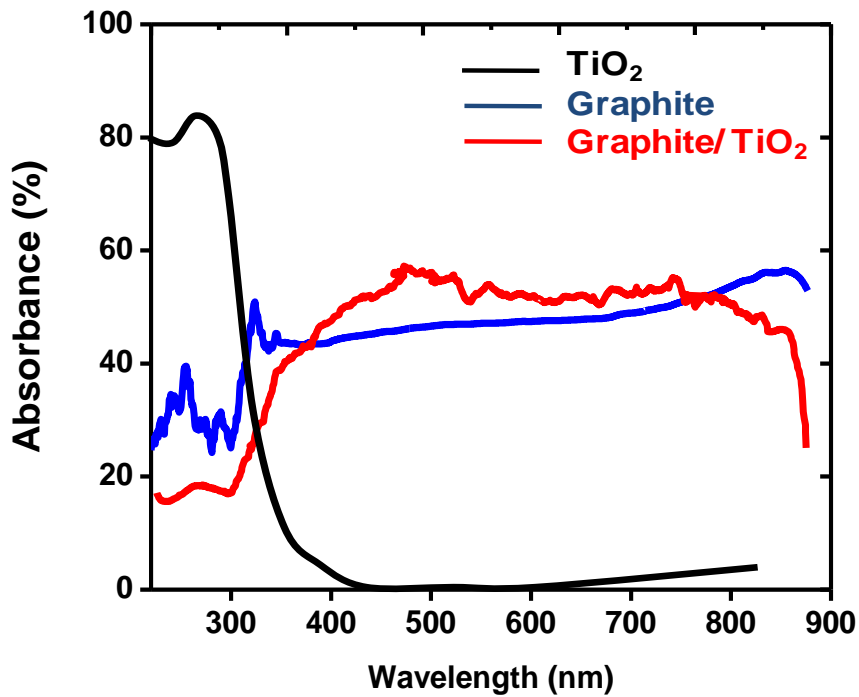

Figure 5. Ultraviolet–visible light characterization of pure $TiO_2$, graphite and graphite/$TiO_2$ composite

Graphite has a wider band of absorption but this does not mean the absorbed photons were used to produce electron–hole pairs. However, we observed a strange phenomenon. Although graphite is a semimetal, its function is similar to that of $TiO_2$ when applied to make solar cells having the structure reported in [1]. Why does a semimetal function similarly to a semiconductor? It remains unclear whether phenomena other than electron–hole pair production play a role in this. Garcia et al. found evidence for the semiconducting behavior of Bernal graphite with a narrow band gap [22], but it is unclear whether the graphite used in the present work exhibited semiconducting behavior.

Further work is required to improve the properties of the presently proposed solar cell. Because the presently reported model's efficiency is lower than that of commercial p–n junction or dye-sensitized solar cells, greatly improving the efficiency of our proposed model remains an important objective. Additionally, the stability of the solar cells (e.g., how long the solar cell can maintain its function before decaying) must be explored intensively. .

## Conclusion

We demonstrated that a design of solar cells having a TiO$_2$/graphite mixture as an active (absorber) material can reach an efficiency of approximately 1.09%, much higher than solar cells using TiO$_2$ only (0.03%) or graphite only (0.12%) as the active material. Owing to its easy and flexible fabrication, and easy scaling up to very large areas, the reported solar cell model is promising for mass application, especially in low-income communities.

## Acknowledgements

This work was supported by a research grant from the Ministry of Research and Higher Education of the Republic of Indonesia (contract no. 310y/l1.C01/PL/2015) and fellowship grand from Indonesia Endowment Fund for Education (LPDP).

## References

[1] S. Saehana, P. Arifin, Khairurrijal, and M. Abdullah, J. Appl. Phys. 111, 123109 (2012).

[2] S. Saehana, E. Yuliza, P. Arifin, Khairurrijal, M. Abdullah, Mater. Sci. Forum 737, 43-53 (2013)

[3] E. Yuliza, S. Saehana, D. Y. Rahman, M. Rosi, Khairurrijal, M. Aabdullah, Mater. Sci. Forum 737, 85-92 (2013)

[4] S. Saehana, R. Prasetyowati, M.I. Hidayat, P. Arifin, and M. Abdullah, Int. J. Basic Appl. Sci. 12, 15 (2011)

[5] M. Rokhmat, E. Wibowo, Sutisna, E. Yuliza, Khairurrijal and M. Abdullah, Adv. Mater. Res. 1112, 245-250 (2015)

[6] B. O'regan and M. Grfitzeli, Nature 353, 737-740 (1991).

[7] N. Serpone and E. Pelizzetti, Photocatalys, Fundamental and Application, Wiley new York (1989)

[8] X.Z.Li and F.B. Li, Environ. Sci. Technol. 35, 2381-2387 (2001)

[9] Y. Wang, H. Cheng, L. Zhang, Y. Hao, J. Ma. B. Xu and W. Li, J. Mol. Cat. A: Chem. 151, 205-216 (2000)

[10] M. Abdullah, I. Nurmawarti, H. Subianto, Khairurrijal, H. Mahfudz, J. Nanosains Nanoteknol. 3, 10-14 (2010)

[11] Y. Yamada and Y. Kanemitsu, Appl. Phys. Lett. 101, 133907 (2012)

[12] A. V. Vorontsov, E.N. Savinov, and Z.S. Jin, J. Photochem. Photobiol. A125, 113-117 (1999)

[13] F.B. Li and X.Z. Li, Chemosphere 48, 1103-1111 (2001)

[14] M.R. Hoffmann, S.T. Martin, W. Choi, and D.W. Bahnemannt, Chem. Rev. 95, 69-96 (1995)

[15] S. Banerjee, J. Gopal, P. Muraleedharan, A.K. Tyagi, and B. Raj, Curr. Sci. 90, No. 10 (2006)

[16] J.Y. Kim, K. Lee, N.E. Coates, D. Moses, T.Q. Nguyen, M. Dante, and A.J. Heeger, Science, 317(5835), 222-225 (2007).

[17] J. Peet, J.Y. Kim, N.E. Coates, W.L. Ma, D. Moses, A.J. Heeger, and G.C. Bazan, Nature Mater. 6, 497-500 (2007)

[18] R. Sengupta, M. Bhattacharya, S. Bandyopadhyay, and A.K. Bhowmick, Prog. Polym. Sci. 36, 638-670 (2011).

[19] J.W. McClure, Phys. Rev. 108, 612 (1957)

[20] J.W. McClure, IBM J. July 1964

[21] P.B. Nair, V.B. Justinvictor, G.P. Daniel, K, Joy and P.V. Thomas, J. Mater. Sci.: Mater. Electron. 24, 2453-2460 (2013).

[22] N. García, P. Esquinazi, J. Barzola-Quiquia, and S. Dusari, New J. Phys. 14, 053015 (2012).